\documentclass[aps,prl,twocolumn,superscriptaddress,floatfix]{revtex4}
\usepackage{amsmath,fancybox}
\usepackage{graphicx}
\usepackage{epsfig,subfigure}

\def\mom{{\rm GeV}/c}
\def\gev{{\rm GeV}/c^2}
\def\mev{{\rm MeV}/c^2}

\begin{document}

%\preprint{}

\title{Exotic Meson Production in the $f_{1}(1285)\pi^{-}$ System observed\\
\boldmath{in the Reaction $\pi^{-} p \rightarrow \eta\pi^{+}\pi^{-}\pi^{-} p$ at $18 \ \mom$}}

% My list of collaborators:
% ====================================
\author{J.~\surname{Kuhn}}
\email[E-mail: ]{kuhnj@ernest.phys.cmu.edu}
\altaffiliation[Present address: ]{Department of Physics, Carnegie Mellon University, Pittsburgh, Pennsylvania 15213}
\affiliation{Department of Physics, Rensselaer Polytechnic Institute, Troy, New York 12180}
\author{G.~S.~\surname{Adams}}
\affiliation{Department of Physics, Rensselaer Polytechnic Institute, Troy, New York 12180}
\author{T.~\surname{Adams}}
\altaffiliation[Present address: ]{Department of Physics, Florida State University, Tallahassee, FL 32306}
\affiliation{Department of Physics, University of Notre Dame, Notre Dame, Indiana 46556}
\author{Z.~\surname{Bar-Yam}}
\affiliation{Department of Physics, University of Massachusetts Dartmouth, North Dartmouth, Massachusetts 02747}
\author{J.~M.~\surname{Bishop}}
\affiliation{Department of Physics, University of Notre Dame, Notre Dame, Indiana 46556}
\author{V.~A.~\surname{Bodyagin}}
\affiliation{Nuclear Physics Institute, Moscow State University, Moscow, Russian Federation 119899}
\author{D.~S.~\surname{Brown}}
\altaffiliation[Present address: ]{Department of Physics, University of Maryland, College Park, MD 20742}
\affiliation{Department of Physics, Northwestern University, Evanston, Illinois 60208}
\author{N.~M.~\surname{Cason}}
\affiliation{Department of Physics, University of Notre Dame, Notre Dame, Indiana 46556}
\author{S.~U.~\surname{Chung}}
\affiliation{Physics Department, Brookhaven National Laboratory, Upton, New York 11973}
\author{J.~P.~\surname{Cummings}}
\affiliation{Department of Physics, Rensselaer Polytechnic Institute, Troy, New York 12180}
\author{K.~\surname{Danyo}}
\affiliation{Physics Department, Brookhaven National Laboratory, Upton, New York 11973}
\author{A.~I.~\surname{Demianov}}
\affiliation{Nuclear Physics Institute, Moscow State University, Moscow, Russian Federation 119899}
\author{S.~P.~\surname{Denisov}}
\affiliation{Institute for High Energy Physics, Protvino, Russian Federation 142284}
\author{V.~\surname{Dorofeev}}
\affiliation{Institute for High Energy Physics, Protvino, Russian Federation 142284}
\author{J.~P.~\surname{Dowd}}
\affiliation{Department of Physics, University of Massachusetts Dartmouth, North Dartmouth, Massachusetts 02747}
\author{P.~\surname{Eugenio}}
\affiliation{Department of Physics, Florida State University, Tallahassee, FL 32306}
\author{X.~L.~\surname{Fan}}
\affiliation{Department of Physics, Northwestern University, Evanston, Illinois 60208}
\author{A.~M.~\surname{Gribushin}}
\affiliation{Nuclear Physics Institute, Moscow State University, Moscow, Russian Federation 119899}
\author{R.~W.~\surname{Hackenburg}}
\affiliation{Physics Department, Brookhaven National Laboratory, Upton, New York 11973}
\author{M.~\surname{Hayek}}
\altaffiliation[Permanent address: ]{Rafael, Haifa, Israel}
\affiliation{Department of Physics, University of Massachusetts Dartmouth, North Dartmouth, Massachusetts 02747}
\author{J.~\surname{Hu}}
\altaffiliation[Present address: ]{TRIUMF, Vancouver, B.C., V6T 2A3, Canada}
\affiliation{Department of Physics, Rensselaer Polytechnic Institute, Troy, New York 12180}
\author{E.~I.~\surname{Ivanov}}
\affiliation{Department of Physics, Idaho State University, Pocatello, ID 83209}
\author{D.~\surname{Joffe}}
\affiliation{Department of Physics, Northwestern University, Evanston, Illinois 60208}
\author{I.~\surname{Kachaev}}
\affiliation{Institute for High Energy Physics, Protvino, Russian Federation 142284}
\author{W.~\surname{Kern}}
\affiliation{Department of Physics, University of Massachusetts Dartmouth, North Dartmouth, Massachusetts 02747}
\author{E.~\surname{King}}
\affiliation{Department of Physics, University of Massachusetts Dartmouth, North Dartmouth, Massachusetts 02747}
\author{O.~L.~\surname{Kodolova}}
\affiliation{Nuclear Physics Institute, Moscow State University, Moscow, Russian Federation 119899}
\author{V.~L.~\surname{Korotkikh}}
\affiliation{Nuclear Physics Institute, Moscow State University, Moscow, Russian Federation 119899}
\author{M.~A.~\surname{Kostin}}
\affiliation{Nuclear Physics Institute, Moscow State University, Moscow, Russian Federation 119899}
\author{V.~V.~\surname{Lipaev}}
\affiliation{Institute for High Energy Physics, Protvino, Russian Federation 142284}
\author{J.~M.~\surname{LoSecco}}
\affiliation{Department of Physics, University of Notre Dame, Notre Dame, Indiana 46556}
\author{M.~\surname{Lu}}
\affiliation{Department of Physics, Rensselaer Polytechnic Institute, Troy, New York 12180}
\author{J.~J.~\surname{Manak}}
\affiliation{Department of Physics, University of Notre Dame, Notre Dame, Indiana 46556}
\author{J.~\surname{Napolitano}}
\affiliation{Department of Physics, Rensselaer Polytechnic Institute, Troy, New York 12180}
\author{M.~\surname{Nozar}}
\altaffiliation[Present address: ]{Thomas Jefferson National Accelerator Facility, Newport News, Virginia 23606}
\affiliation{Department of Physics, Rensselaer Polytechnic Institute, Troy, New York 12180}
\author{C.~\surname{Olchanski}}
\altaffiliation[Present address: ]{TRIUMF, Vancouver, B.C., V6T 2A3, Canada}
\affiliation{Physics Department, Brookhaven National Laboratory, Upton, New York 11973}
\author{A.~I.~\surname{Ostrovidov}}
\affiliation{Department of Physics, Florida State University, Tallahassee, FL 32306}
\author{T.~K.~\surname{Pedlar}}
\altaffiliation[Present address: ]{Laboratory for Nuclear Studies, Cornell University, Ithaca, NY 14853}
\affiliation{Department of Physics, Northwestern University, Evanston, Illinois 60208}
\author{A.~V.~\surname{Popov}}
\affiliation{Institute for High Energy Physics, Protvino, Russian Federation 142284}
\author{D.~I.~\surname{Ryabchikov}}
\affiliation{Institute for High Energy Physics, Protvino, Russian Federation 142284}
\author{L.~I.~\surname{Sarycheva}}
\affiliation{Nuclear Physics Institute, Moscow State University, Moscow, Russian Federation 119899}
\author{K.~K.~\surname{Seth}}
\affiliation{Department of Physics, Northwestern University, Evanston, Illinois 60208}
\author{N.~\surname{Shenhav}}
\altaffiliation[Permanent address: ]{Rafael, Haifa, Israel}
\affiliation{Department of Physics, University of Massachusetts Dartmouth, North Dartmouth, Massachusetts 02747}
\author{X.~\surname{Shen}}
\altaffiliation[Permanent address: ]{Institute of High Energy Physics, Bejing, China}
\affiliation{Department of Physics, Northwestern University, Evanston, Illinois 60208}
\affiliation{Thomas Jefferson National Accelerator Facility, Newport News, Virginia 23606}
\author{W.~D.~\surname{Shephard}}
\affiliation{Department of Physics, University of Notre Dame, Notre Dame, Indiana 46556}
\author{N.~B.~\surname{Sinev}}
\affiliation{Nuclear Physics Institute, Moscow State University, Moscow, Russian Federation 119899}
\author{D.~L.~\surname{Stienike}}
\affiliation{Department of Physics, University of Notre Dame, Notre Dame, Indiana 46556}
\author{J.~S.~\surname{Suh}}
\altaffiliation[Present address: ]{Department of Physics, Kyungpook National University, Daegu, Korea}
\affiliation{Physics Department, Brookhaven National Laboratory, Upton, New York 11973}
\author{S.~A.~\surname{Taegar}}
\affiliation{Department of Physics, University of Notre Dame, Notre Dame, Indiana 46556}
\author{A.~\surname{Tomaradze}}
\affiliation{Department of Physics, Northwestern University, Evanston, Illinois 60208}
\author{I.~N.~\surname{Vardanyan}}
\affiliation{Nuclear Physics Institute, Moscow State University, Moscow, Russian Federation 119899}
\author{D.~P.~\surname{Weygand}}
\affiliation{Thomas Jefferson National Accelerator Facility, Newport News, Virginia 23606}
\author{D.~B.~\surname{White}}
\affiliation{Department of Physics, Rensselaer Polytechnic Institute, Troy, New York 12180}
\author{H.~J.~\surname{Willutzki}}
\altaffiliation{Deceased, Dec. 2001}
\affiliation{Physics Department, Brookhaven National Laboratory, Upton, New York 11973}
\author{M.~\surname{Witkowski}}
\affiliation{Department of Physics, Rensselaer Polytechnic Institute, Troy, New York 12180}
\author{A.~A.~\surname{Yershov}}
\affiliation{Nuclear Physics Institute, Moscow State University, Moscow, Russian Federation 119899}
% =====  End of collaborator's list  =====

% Collaboration name
% ==================
\collaboration{The E852 collaboration}
% =====  End of collaboration name  =====

\date{\today}

% Abstract
% ========
\begin{abstract}
This letter reports results from the partial wave analysis of the $\pi^{-}\pi^{-}\pi^{+}\eta$ final state in $\pi^{-}p$ collisions at $18 \ \mom$.  Strong evidence is observed for production of two mesons with exotic quantum numbers of spin, parity and charge conjugation, $J^{PC} = 1^{-+}$ in the decay channel $f_{1}(1285)\pi^{-}$. The mass $M = 1709 \pm 24 \pm 41 \ \mev$ and width $\Gamma = 403 \pm 80 \pm 115 \ \mev$ of the first state are consistent with the parameters of the previously observed $\pi_{1}(1600)$. The second resonance with mass $M = 2001 \pm 30 \pm 92 \ \mev$ and width $\Gamma = 333 \pm 52 \pm 49 \ \mev$ agrees very well with predictions from theoretical models.  In addition, the presence of $\pi_{2}(1900)$ is confirmed with mass $M = 2003 \pm 88 \pm 148 \ \mev$ and width $\Gamma = 306 \pm 132 \pm 121 \ \mev$ and a new state, $a_{1}(2096)$, is observed with mass $M = 2096 \pm 17 \pm 121 \ \mev$ and width $\Gamma = 451 \pm 41 \pm 81 \ \mev$.  The decay properties of these last two states are consistent with flux tube model predictions for hybrid mesons with non-exotic quantum numbers.

\end{abstract}

%\keywords{}

\maketitle

% Body of the paper
% =================

% Introduction
% ============
\section{Introduction}

States outside the constituent quark model have been hypothesized to exist almost since the introduction of color~\cite{QuarkModels1,QuarkModels2,QuarkModels3,QuarkModels4}.  Hybrid mesons, $q\bar{q}$ states with an admixture of gluons, and glueballs, states with no quark content, rely on the self interaction property of gluons due to their color charge.  Looking for glueballs would be the most obvious way to find evidence for states with constituent gluons; however, the search is hindered by the fact that these states may significantly mix with regular $q\bar{q}$-mesons in the region where the lightest are predicted to occur.  As such, they may not be observable as pure states and disentangling the observed spectra may be a very difficult task.  Instead, hybrid mesons ($q\bar{q}g^{n}$) may be a better place to search for evidence of resonances outside the constituent quark model, especially since the lightest of theses states are predicted to have exotic quantum numbers of spin, parity, and charge conjugation, $J^{PC}$, that is, combinations that are unattainable by regular $q\bar{q}$-mesons.

Several candidates for $J^{PC}=1^{-+}$ exotic states have been reported in the last few years.  In the $\eta\pi^{-}$ decay channel strong evidence was discovered for a state at around $1.4 \ \gev$ by the BNL experiment E852~\cite{EtaPi}.  This was later confirmed in an independent analysis by the Crystal Barrel collaboration~\cite{EtaPiCB}.  A second $J^{PC}=1^{-+}$ state at $1.6 \ \gev$ has been observed in two decay modes, $\rho^{0}\pi^{-}$~\cite{PiPiPi} and $\eta'\pi^{-}$~\cite{EtaPrPi}.  The interpretation of these states as hybrids is still unclear since the flux tube model predicts that the decay of a $J^{PC}=1^{-+}$ hybrid into two $S$-wave mesons should be highly suppressed as compared to decays into a $P$- and an $S$-wave meson~\cite{HybridPhen,ClosePage}, and their masses are lower than model predictions of $1.8-2.0 \ \gev$~\cite{QuarkModels4,hybLGT1,hybLGT2}.  The non-$q\bar{q}$ quantum numbers can also be explained by formation of a four quark state ($q\bar{q}q\bar{q}$).

An analysis of the $K^{+}\bar{K}^{0}\pi^{-}\pi^{-}$ final state by BNL experiment E818~\cite{Lee} reported a broad structure in the $J^{PC}=1^{-+}$ wave in the region from $1.6$ to $2.2 \ \gev$, which suggested two objects, one at $1.7 \ \gev$ and the other around $2.0 \ \gev$. The lower mass object had a substantial coupling to the final state $\eta(1295)\pi^{-}$, whereas the one at higher mass was dominated by $f_{1}(1285)\pi^{-}$. However, due to lack of statistics a firm conclusion on the resonant behavior of these two states was not possible.

The main objective of the analysis presented here was to look for the decay $X^{-}(J^{PC}=1^{-+}) \rightarrow f_{1}(1285)\pi^{-}$, which is predicted to be one of the strongest decay modes of the lightest exotic hybrid~\cite{ClosePage}. The $f_{1}(1285)$ was detected in its decay mode to $\pi^{-}\pi^{+}\eta$.
An earlier analysis of this decay channel with data from the 1994 data taking period of E852 showed agreement with the data from E818, but lack of statistics made it difficult to resolve states in the spectra~\cite{ToddThesis}.

% Data selection
% ==============
\section{Experimental Setup and Data selection}

Data for this analysis were taken during the 1995 running of E852 at the Multi-Particle Spectrometer.  The Alternating Gradient Synchrotron provided an $18 \ {\rm GeV}/c \ \pi^{-}$ beam which was incident upon a 30-cm long liquid hydrogen target.  A detailed description of the experiment can be found elsewhere~\cite{EtaPi,ThesisJK}.  The analysis presented here is based on the $p\pi^{+}\pi^{-}\pi^{-}\gamma\gamma$ final state which is a subset of the approximately $10^{9}$ triggers collected in this run period.  The trigger for this data set required a large-angle charged track in the cylindrical drift chamber surrounding the target, three charged tracks in downstream tracking chambers and energy deposited in the Lead Glass Detector (LGD).

\begin{figure}[here,top]
\centering
\includegraphics[width=8.65cm]{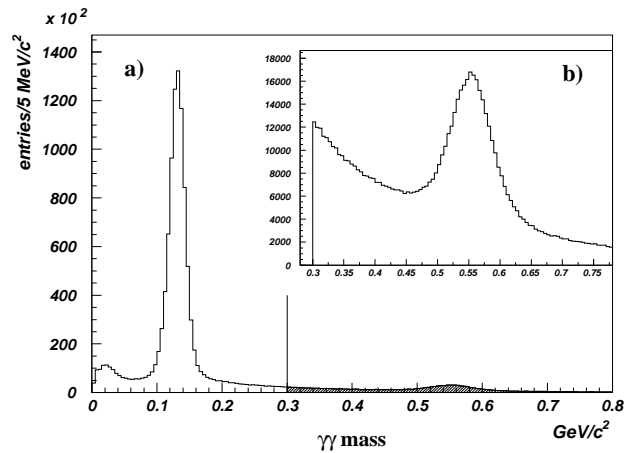}
\caption{\label{fig:ggmass} (a) $\gamma\gamma$ mass after selecting 2 photons in the LGD for a sample of the total data set. The vertical line indicates the cut to select $\eta \rightarrow \gamma\gamma$ events. (b) $\gamma\gamma$ mass after additional fiducial and geometry cuts.}
\end{figure}

In the first stage of the analysis charged tracks were reconstructed and a requirement on the number of photon candidates in the lead glass detector array was imposed.  This reduced the total initial data set from $265 \times 10^{6}$ to about $10.5 \times 10^{6}$ events.  As can be seen from the $\gamma\gamma$ mass distribution (Fig.~\ref{fig:ggmass}a) the spectrum at this stage is dominated by events containing $\pi^{0} \rightarrow \gamma\gamma$ decays.  The two-photon mass was therefore restricted to values greater than $300 \ \mev$, indicated by the vertical line in Fig.~\ref{fig:ggmass}a, in order to select events containing $\eta(547)$ in its decay mode to $\gamma\gamma$.  At this stage a fiducial cut on the target volume and on the LGD were performed as well, reducing the data set to approximately $750 \ 000$ events.  The $\gamma\gamma$ mass after these cuts presents a clear signal for the $\eta$ with a signal-to-background ratio of approximately $1:1$ in the mass region from $0.45$ to $0.65 \ \gev$ (see Fig.~\ref{fig:ggmass}b).  The mass of the recoil particle and the $\gamma\gamma$ mass were constrained to the mass of the proton and to the mass of the $\eta$ in a kinematic fit~\cite{SQUAWprog}.  The confidence-level from the fit was required to be greater than 10\%.  

Further improvement in the sample purity was achieved by comparing the measured $\phi$ angle of the recoil track to the proton angle obtained from the fit. Events were removed if the difference between these two angles was greater than $20^{\circ}$.  A small elliptical region corresponding to the size of the beam envelope passing through the chambers was defined. Events with tracks passing through this region where the chambers became inefficient were removed from the sample.

\begin{figure}[here,top]
\centering
\includegraphics[width=6.00cm]{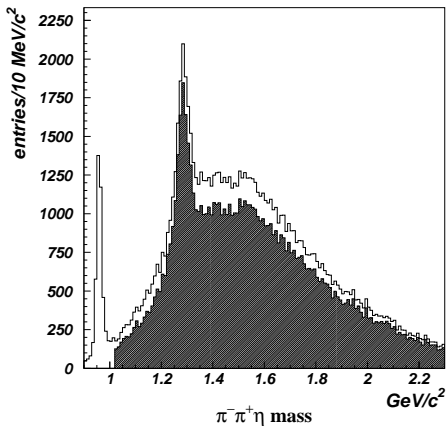}
\caption{\label{fig:ppemass} $\pi^{-}\pi^{+}\eta$ mass distribution before (unhatched) and after (hatched) rejecting events that contain $\eta'(958) \rightarrow \pi^{-}\pi^{+}\eta$. Note that there are two entries for each event.}
\end{figure}

\begin{figure}[here,top]
\centering
\mbox{\subfigure{\epsfig{figure=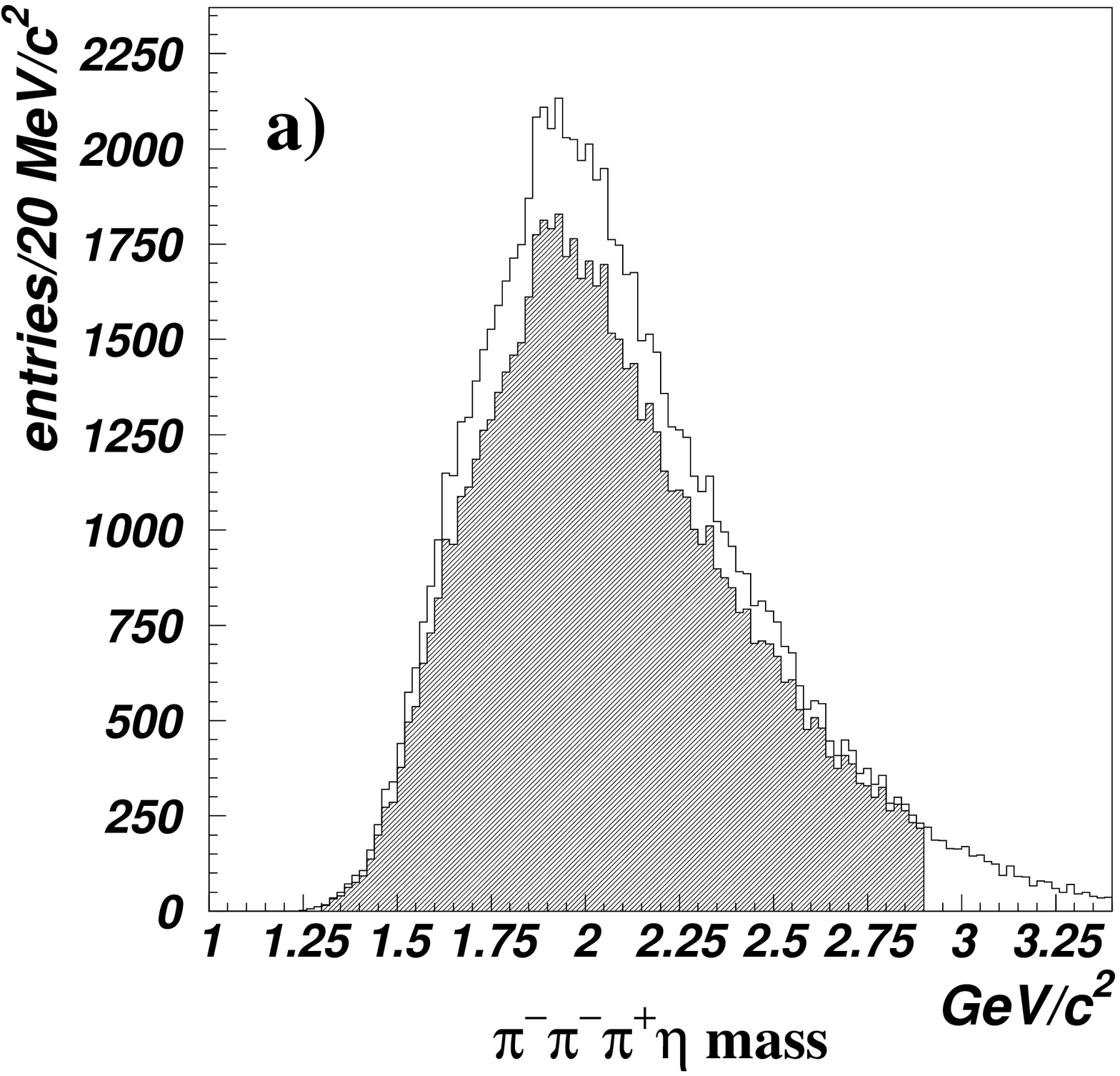,width=4.0cm}}
\subfigure{\epsfig{figure=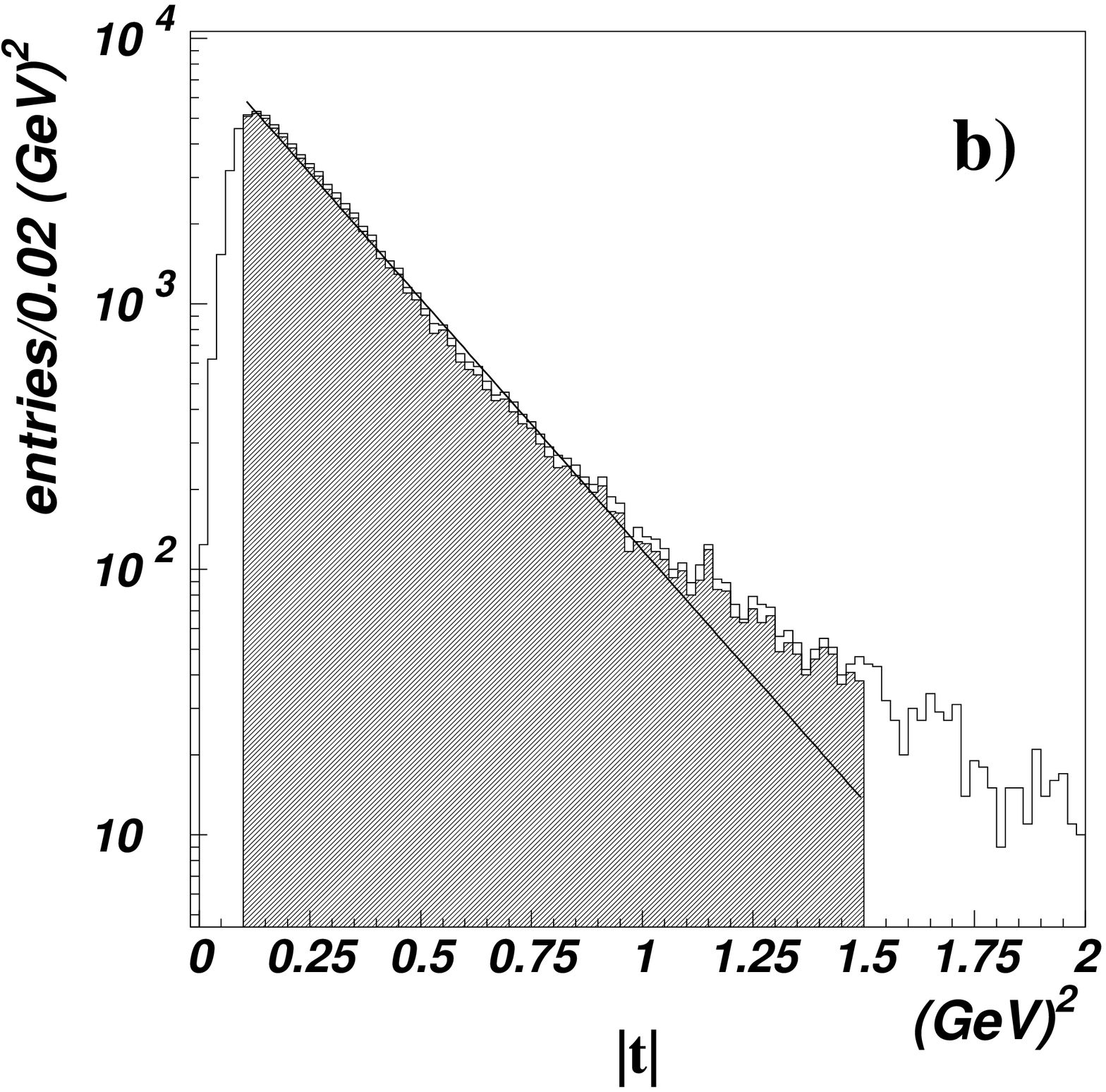,width=4.0cm}}}
\caption{\label{fig:mesmass} (a) $3\pi\eta$ mass for the final data set. (b) Four momentum transfer distribution for the final data set. The distribution is fitted to the function $f(t) \propto e^{b|t|}$. The hatched distributions indicate events that are passed to the PWA.}
\end{figure}

The $\pi^{-}\pi^{+}\eta$ invariant mass (Fig.~\ref{fig:ppemass}) clearly shows the $\eta'(958)$ and the $f_{1}(1285)$/$\eta(1295)$.  The hatched histogram shows the event sample after rejecting events where either of the two possible combinations of $\pi^{-}\pi^{+}\eta$ were consistent with the mass of the $\eta'(958)$. This cut significantly reduces the number of waves required by the partial wave analysis.
Of the $82 \ 645$ events in the final data set, $68 \ 900$ were passed to the partial wave analysis (PWA) by restricting the $\pi^{-}\pi^{-}\pi^{+}\eta$ ($3\pi\eta$) mass to values between $1.3$ and $2.9 \ \gev$ and the four momentum transfer, $t$, to the range $-1.5 \leq t \leq -0.1 \ ({\rm GeV}/c)^{2}$.  Between $-1.5 \ ({\rm GeV}/c)^{2}$ and $-0.1 \ ({\rm GeV}/c)^{2}$ the $t$-distribution can be fitted to the function $f(t) \propto e^{b|t|}$ with $b = -4.36$, consistent with peripheral production of mesons.  This range in $t$ was chosen because the detector has low acceptance at $t$ near zero, and a change in slope below $-1.0 \ ({\rm GeV}/c)^{2}$ may indicate a new production mechanism (see Fig.~\ref{fig:mesmass}b).  The PWA was performed on the data set shown as the hatched histograms in Fig.~\ref{fig:mesmass}.

% GAMMA-GAMMA MASS PLOTS

% 2PI-ETA MASS PLOT

% FINAL DATA SET PLOTS (3pieta mass and t-dist)

% Partial Wave Analysis
% =====================

\section{Partial Wave Analysis}

\subsection{PWA - Description of the method}

Since there are many wide and overlapping states with different $J^{PC}$ that contribute to the $3\pi\eta$ final state in the mass range of interest, the extraction of these states is facilitated by performing a partial wave analysis, where not only the mass, but also the angular distributions are taken into account.

In the PWA program~\cite{PWAprog} the isobar model~\cite{Hernon} is used to describe the decay of each event.  In this model the $X^{-}$ system decays into the $3\pi\eta$ final state via successive two-body decays, which can occur via two paths
\begin{equation}\label{eq:Decay1}
\begin{split}
X^{-} \rightarrow &I_{1} + B_{1} \\
&I_{1} \rightarrow I_{2} + B_{2} \\
& \hspace{8.7mm} I_{2} \rightarrow B_{3} + B_{4}
\end{split}
\end{equation}
\begin{equation}\label{eq:Decay2}
\begin{split}
X^{-} \rightarrow &I_{1} + I_{2} \\
&I_{1,2} \rightarrow B_{1,3} + B_{2,4}
\end{split}
\end{equation}
The decay products are classified into two categories, {\em isobars} ($I_{n}$), which themselves decay further, and {\em bachelors} ($B_{n}$), which are the final state particles and are considered stable.

The total intensity of the $3\pi\eta$ system is a superposition of both interfering and non-interfering sets of states produced in the reaction, and can be written as
\begin{equation}\label{eq:PWA_Int}
I(\tau) = \sum_{\epsilon,k} \biggl| \sum_{\alpha} {}^{\epsilon}V_{k\alpha} \ {}^{\epsilon}A_{\alpha}(\tau) \biggr|^{2}
\end{equation}
where ${}^{\epsilon}V_{k\alpha}$ and ${}^{\epsilon}A_{\alpha}(\tau)$ are the production and decay amplitudes, calculated in the helicity basis~\cite{SUCyr}. The variable $\tau$ labels the eight kinematic variables describing the $3\pi\eta$ events. The variable $k$ corresponds to the two possible orientations of the proton spin, thus restricting the rank of the spin-density matrix to no more than two and $\epsilon$ is the reflectivity of the state $X^{-}$. The decay amplitudes are calculated in the reflectivity basis since for a $\pi$ beam the reflectivity of the produced state coincides with the naturality of the exchanged particle and parity conservation constraints break up the spin-density matrix into two block diagonal sub-matrices~\cite{SUCTrue}.
Each wave entering the PWA is described by a set of quantum numbers, $\alpha$
\begin{equation}\label{eq:PWA_Vars}
\alpha = \{J^{PC} \ L \ S \ l_{1} \ s_{1} \ l_{2} \ s_{2} \ m \ (M_{1},\Gamma_{1}) \ (M_{2},\Gamma_{2})\}
\end{equation}
where $L$ and $l_{n}$ indicate the orbital angular momenta between the two-body decay products, $S$ and $s_{n}$ their spins, $m$ the projection of the spin $J$ onto the $z$-axis and $M_{1,2}$ and $\Gamma_{1,2}$ the masses and widths of the isobars in the reaction. In this letter we concentrate on results from states decaying via the primary decay chain (\ref{eq:Decay1}). For simplicity these waves will be labeled as $J^{PC} m^{\epsilon} [I_{1} \ B_{1}] L$. 

The quantities ${}^{\epsilon}V_{k\alpha}$ are independent of $\tau$ and are determined from an extended maximum-likelihood fit. The experimental acceptance was determined by means of a Monte Carlo simulation and then incorporated into the normalization for each wave~\cite{PWAprog}.

Partial waves were restricted to $J,L \leq 4$ since high spin waves are believed to be of little significance in the low mass region under consideration. Since the proton helicities of the initial and final state differ at most by 1, $m$ was restricted to $|m| \leq 1$~\cite{EtaPi,ThesisJK}. 

The decay of the isobars $f_{1}(1285)$ and $\eta(1295)$ into $\pi^{+}\pi^{-}\eta$ can occur via the following three modes
\begin{equation}\label{eq:F1Decays}
f_{1}/\eta \rightarrow a_{0}^{+}\pi^{-} \ \ \ f_{1}/\eta \rightarrow a_{0}^{-}\pi^{+} \ \ \ f_{1}/\eta \rightarrow \sigma\eta
\end{equation}
where $\sigma$ is the $(\pi\pi)_S$-wave interaction and $a_{0}^{\pm} \rightarrow \eta\pi^{\pm}$. In order to reduce the number of parameters entering the fit, amplitudes containing the first two of these decays were added together using isospin conservation
\begin{equation}\label{eq:IsobarAddF1Pi}
\begin{split}
A(f_{1}/\eta \rightarrow a_{0}\pi) = &\frac{1}{\sqrt{3}} \bigl[ A(f_{1}/\eta \rightarrow a_{0}^{-}\pi^{+})\\
&+ A(f_{1}/\eta \rightarrow a_{0}^{+}\pi^{-}) \bigr]
\end{split}
\end{equation}
We verified from the fits that $A(f_{1}/\eta \rightarrow a_{0}^{+}\pi^{-}) = A(f_{1}/\eta \rightarrow a_{0}^{-}\pi^{+})$ before constraining the waves in later fits.  In a similar fashion the decays of the $a_{1}^{-}(1260)$ into $\rho\pi^{-}$, which can occur with relative orbital angular momentum $l=0$ ($S$-wave) or $l=2$ ($D$-wave), were combined into one wave, using
\begin{equation}\label{eq:IsobarAddA1Eta}
A(a_{1}^{-} \rightarrow \rho^{0}\pi^{-}) = A(a_{1}^{-} \rightarrow (\rho^{0}\pi^{-})_{S}) \bigl[ 1+ R e^{-i\Delta\phi} \bigr]
\end{equation}
where the ratio $R = |D| / |S|$ and the phase difference $\Delta\phi = \phi_{S}-\phi_{D}$ were determined from several different waves in the fit. They were found to be 
\begin{equation}
R = 0.24 \pm 0.02\ \ \ \Delta\phi = -2.3 \pm 0.14
\end{equation}
in agreement with an earlier measurement~\cite{PiPiPi}.  The parametrization by Au, Morgan, and Pennington~\cite{AMP} was used to describe the $\sigma$. For the $f_{1}(1285)$ and the $\eta(1295)$ isobars, the experimental decay widths were used, taking into account the resolution of the apparatus. For other states the published values were employed~\cite{PDG}. 

Every fit contained a non-interfering background term with isotropic angular dependence in order to account for non-resonant background and for small waves which were omitted from the fit. Numerous fits with varying wave sets were performed in order to achieve a good description of the various mass and angular distributions. The final wave set was composed of 53 partial waves (complex amplitudes) for each of the two ranks plus the above mentioned background wave, which amounted to a fit with $(53 \times 4 + 1) = 213$ parameters.  The list of waves used is shown in Table~\ref{TableWaves}.

\begin{figure}
\includegraphics[height=8.65cm,angle=-90]{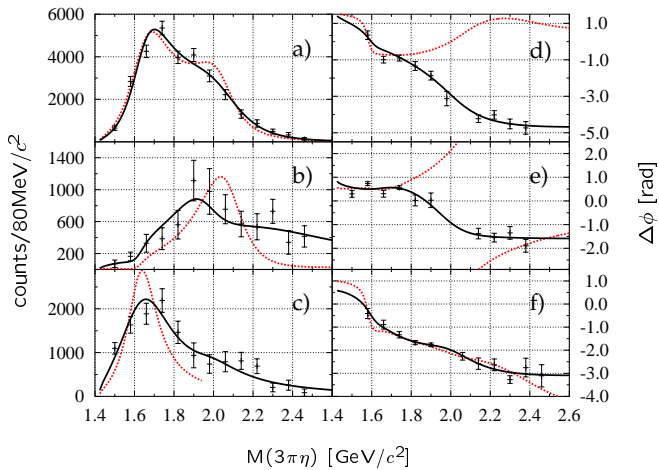}
\caption{\label{fig:PWAresults}PWA results: $f_{1}(1285)\pi^{-}$ intensity distributions (a) $1^{++}0^{+}f_{1}\pi^{-}P$, (b) $2^{-+}0^{+}f_{1}\pi^{-}D$, (c) $1^{-+}1^{+}f_{1}\pi^{-}S$ and phase difference distributions (d) $\phi(1^{-+})-\phi(2^{-+})$, (e) $\phi(1^{-+})-\phi(1^{++})$, (f) $\phi(1^{++})-\phi(2^{-+})$. The results from a least squares fit are overlaid as the solid line (two poles in the $1^{-+}f_{1}\pi$ wave) and the dashed line (one pole in the $1^{-+}f_{1}\pi$ wave).}
\end{figure}

\begin{figure}[here,top]
\centering
\mbox{\subfigure{\epsfig{figure=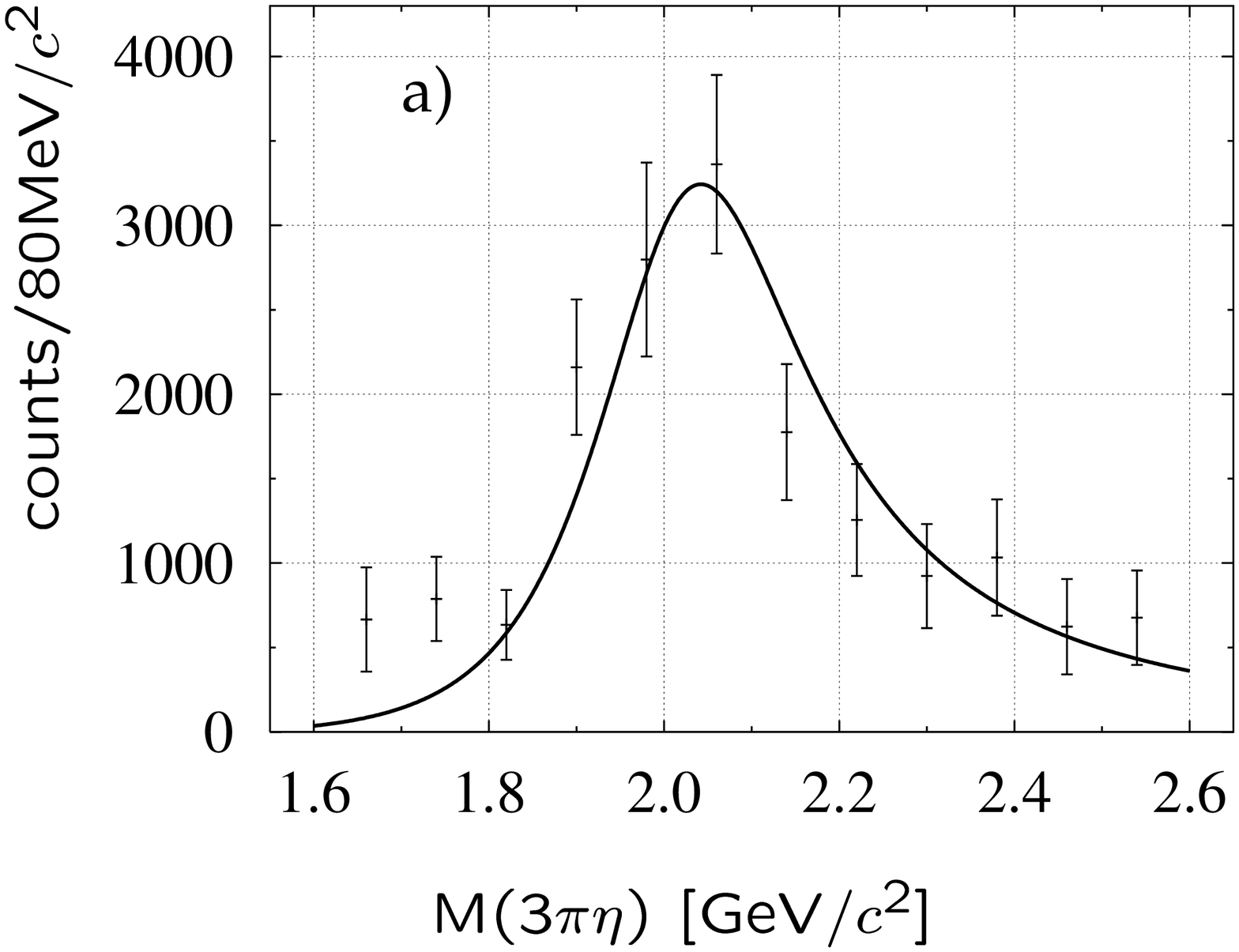,height=4.2cm,angle=-90}}
\subfigure{\epsfig{figure=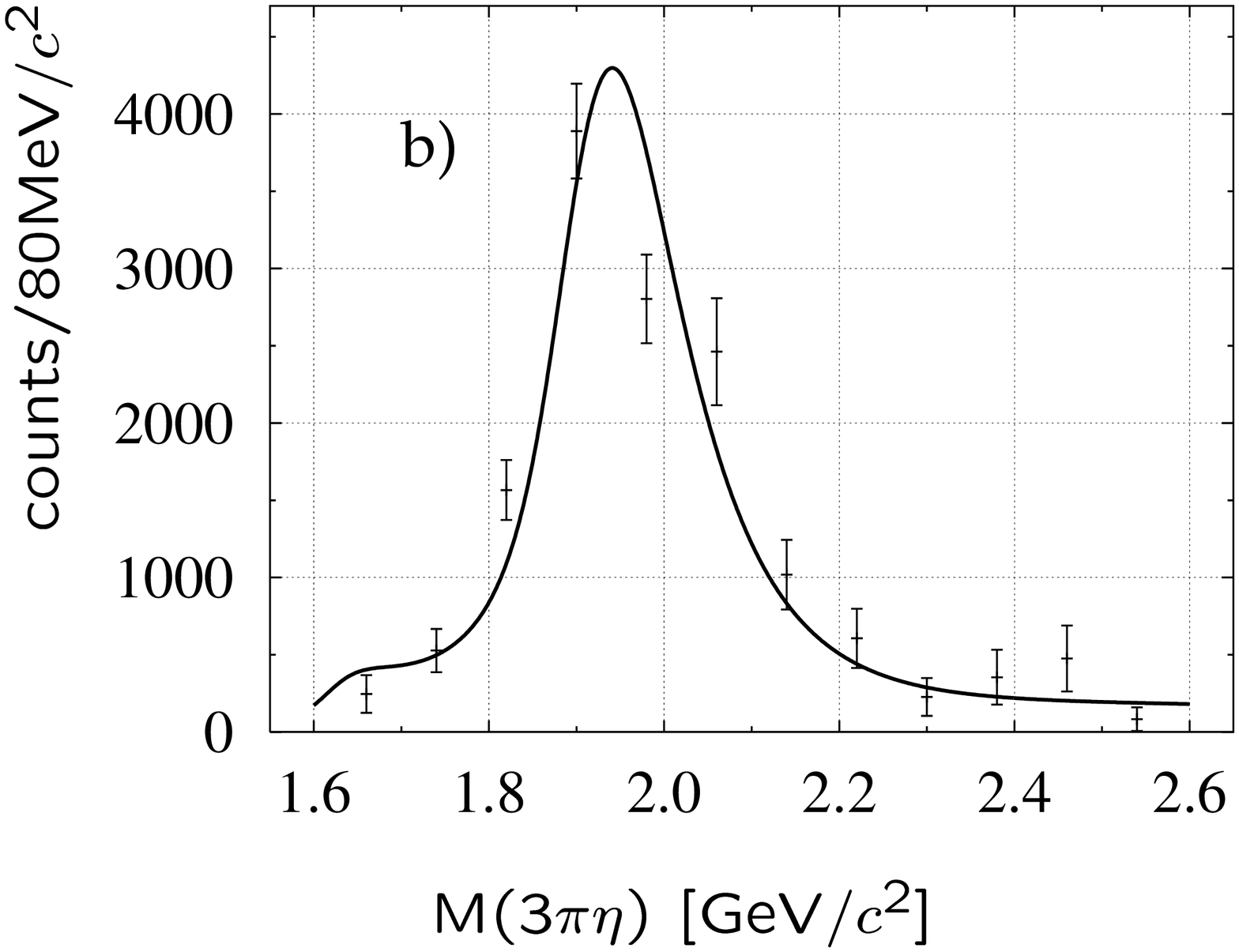,height=4.2cm,angle=-90}}}
\caption{\label{fig:a12eta}PWA results: Intensity distributions for the (a) $1^{++}0^{+} a_{1}^{-}(1260)\eta P$ and (b) $2^{-+}0^{+} a_{2}^{-}(1320)\eta S$ waves. The solid line indicates the result from a fit to Breit-Wigner distributions to determine branching ratios for these waves.}
\end{figure}

The spectrum is dominated by positive reflectivity waves, and the background wave contains approximately $1/3$ of the total strength.  Figs.~\ref{fig:PWAresults} and~\ref{fig:a12eta}  show the three most important $f_{1}\pi^{-}$ waves and the two largest non-$f_{1}\pi^{-}$ waves as the points with the error bars.  Further details on the PWA can be found in reference~\cite{ThesisJK}.  

\begin{table}%[H] add [H] placement to break table across pages
\caption{\label{TableWaves} List of waves used in the final PWA fit.}
\begin{ruledtabular}
\begin{tabular}{l l c c c}
$J^{PC}m^{\epsilon}$  &  primary decay  &  $L$  &  $S$  &  \# of waves \\ \hline
$0^{-+}0^{-}$  &  $\eta(1295)\pi^{-}$  &  0  &  0  &  2\\
$0^{-+}0^{-}$  &  $a_{0}^{-}(980)\sigma$  &  1  &  0  &  1\\
$2^{++}0^{-}$  &  $a_{2}^{-}(1320)\sigma$  &  0  &  2  &  1\\
$2^{++}0^{-}$  &  $a_{2}^{-}(1320)\rho$  &  1,3  &  1,2,3  &  6\\
$1^{-+}1^{+}$  &  $a_{0}^{-}(980)\rho$  &  0  &  1  &  1\\
$1^{-+}1^{+}$  &  $a_{1}^{-}(1260)\eta$  &  0  &  1  &  2\\
$1^{-+}1^{+}$  &  $f_{1}(1285)\pi^{-}$  &  0  &  1  &  2\\
$1^{-+}1^{+}$  &  $\rho'(1460)\pi^{-}$  &  1  &  1  &  1\\
$1^{++}0^{+}$  &  $a_{0}^{-}(980)\rho$  &  1  &  1  &  1\\
$1^{++}0^{+}$  &  $a_{1}^{-}(1260)\eta$  &  1  &  1  &  2\\
$1^{++}0^{+}$  &  $f_{1}(1285)\pi^{-}$  &  1  &  1  &  2\\
$1^{++}0^{+}$  &  $a_{2}^{-}(1320)\eta$  &  1  &  2  &  1\\
$1^{++}0^{+}$  &  $\rho'(1460)\pi^{-}$  &  0,2  &  1  &  2\\
$1^{++}0^{+}$  &  $\rho_{3}(1690)\pi^{-}$  &  2  &  3  &  1\\
$2^{-+}0^{+}$  &  $a_{2}^{-}(1320)\eta$  &  0  &  2  &  1\\
$2^{-+}0^{+}$  &  $\rho'(1460)\pi^{-}$  &  1  &  1  &  1\\
$2^{-+}0^{+}$  &  $a_{1}^{-}(1260)\eta$  &  2  &  1  &  2\\
$2^{-+}0^{+}$  &  $f_{1}(1285)\pi^{-}$  &  2  &  1  &  2\\
$2^{++}1^{+}$  &  $\pi_{2}^{-}(1670)\eta$  &  0  &  2  &  2\\
$2^{++}1^{+}$  &  $a_{2}^{-}(1320)\rho$  &  1  &  1,2,3  &  3\\
$2^{++}1^{+}$  &  $a_{2}^{-}(1320)\eta$  &  1  &  2  &  1\\
$3^{++}0^{+}$  &  $a_{2}^{-}(1320)\eta$  &  1  &  2  &  1\\
$3^{++}0,1^{+}$  &  $a_{2}^{-}(1320)\rho$  &  1  &  2,3  &  4\\
$3^{++}0^{+}$  &  $a_{1}^{-}(1260)\eta$  &  3  &  1  &  2\\
$4^{++}1^{+}$  &  $a_{2}^{-}(1320)\rho$  &  1  &  3  &  1\\
$4^{++}1^{+}$  &  $a_{2}^{-}(1320)\rho$  &  3  &  1,2,3  &  3\\
$4^{++}1^{+}$  &  $a_{1}^{-}(1260)\eta$  &  3  &  1  &  2\\
$4^{++}1^{+}$  &  $a_{2}^{-}(1320)\eta$  &  1  &  2  &  1\\
$4^{++}1^{+}$  &  $\pi^{-}(1800)\eta$  &  4  &  0  &  2\\ \hline
\multicolumn{4}{c}{Background}  &  1\\
\end{tabular}
\end{ruledtabular}
\end{table}

The PWA intensity distributions and phase differences for the waves shown in Fig.~\ref{fig:PWAresults} were fitted by a least squares minimization to linear combinations of relativistic Breit-Wigner poles with mass dependent widths and Blatt-Weisskopf barrier factors~\cite{EtaPi}.  The phase difference between two waves $\alpha$ and $\alpha'$ 
\begin{equation}
{}^{\epsilon}\phi_{\alpha\alpha'} = \arg \biggl( \sum_{k} {}^{\epsilon}V_{k\alpha} {}^{\epsilon}V_{k\alpha'}^{*} \biggr)
\end{equation}
is only well-defined if they are produced coherently in the two spin orientations $k$. Therefore, a small number of mass bins were excluded from the fit if the coherence 
\begin{equation}
{}^{\epsilon}C_{\alpha\alpha'} = \frac{\bigl| \sum_{k} {}^{\epsilon}V_{k\alpha} {}^{\epsilon}V_{k\alpha'}^{*} \bigr|}{\sqrt{ (\sum_{k}|{}^{\epsilon}V_{k\alpha}|^{2})(\sum_{k}|{}^{\epsilon}V_{k\alpha'}|^{2}) } }
\end{equation}
dropped below $60\%$.  Since the barrier factors are indeterminate below the $f_{1}\pi^{-}$ threshold, the distributions were only fitted for $M_{3\pi\eta} > m(f_{1}) + m(\pi)$.  In the fit the masses and widths of the previously observed $\pi_{2}(1670)$ and $a_{1}(1700)$ were fixed at the values taken from reference~\cite{PiPiPi}. 

The best fit was achieved with 2 poles each in the $1^{++}0^{+}f_{1}\pi^{-} P$ and $1^{-+}1^{+}f_{1}\pi^{-} S$ waves and three poles in the $2^{-+}0^{+}f_{1}\pi^{-} D$ wave. The resonance parameters for this fit are listed in Table~\ref{TableResults} and an overlay with the results from the PWA fit is shown in Fig.~\ref{fig:PWAresults} as the black solid line.  The quoted widths are the fitted values and do not take experimental resolution into account.  The fit has a chi-squared per degree-of-freedom $\chi^{2}_{\nu} = 1.5$ with 47 degrees-of-freedom.  Negligible variations in the results were observed if the Chung parametrization for the mass dependent width~\cite{SUC_MDwidth} was used.  

A fit with only one pole in the $1^{-+}1^{+}f_{1}\pi^{-} S$ wave was tried as well and the results are shown as the dashed lines in Fig.~\ref{fig:PWAresults}.  For this fit the $1^{-+}1^{+}f_{1}\pi$ intensity was only fitted up to $1.94 \ \gev$, while all the other intensities and phase differences were fitted up to $2.6 \ \gev$.  The dramatic increase in $\chi^{2}_{\nu}$ to 8.3 with 46 degrees-of-freedom, due primarily to the drastic change in behavior of the phase differences of the $J^{PC}=1^{-+}$ wave with respect to the other two waves (see Figs.~\ref{fig:PWAresults}d and~\ref{fig:PWAresults}e) , strongly supports the existence of a second exotic state at higher mass.  The systematic errors on the masses and widths listed in Table~\ref{TableResults} were determined by fitting results from PWA fits with different wave sets and varying the parameters of the $\pi_{2}(1670)$ and $a_{1}(1700)$ within the error bars determined in reference~\cite{PiPiPi}.

% PLOT WITH PWA-RESULTS 

% PLOT WITH A1ETA AND A2ETA FITS 

% TABLE WITH MDA-RESULTS 
\begin{table}%[H] add [H] placement to break table across pages
\caption{\label{TableResults} Results of the mass-dependent fit.}
\begin{ruledtabular}
\begin{tabular}{c c c}
Wave & Mass $[\mev]$ & $\Gamma$ $[\mev]$ \\ \hline
  &  $1714$ (fixed)  &  $308$ (fixed) \\ 
\raisebox{2.5mm}[2mm][0mm]{$1^{++}0^{+}f_{1}\pi P$}  &  $2096 \pm 17 \pm 121$  &  $451 \pm 41 \pm 81$ \\ \hline
  &  $1676$ (fixed)  &  $254$ (fixed) \\ 
$2^{-+}0^{+}f_{1}\pi D$  &  $2003 \pm 88 \pm 148$  &  $306 \pm 132 \pm 121$ \\
  &  $2460 \pm 328 \pm 263$  &  $1540 \pm 1214\pm 718$\\ \hline
  &  $1709 \pm 24 \pm 41$  &  $403 \pm 80 \pm 115$\\ 
\raisebox{2.5mm}[2mm][0mm]{$1^{-+}1^{+}f_{1}\pi S$}  &  $2001 \pm 30 \pm 92$  &  $333 \pm 52 \pm 49$ \\
\end{tabular}
\end{ruledtabular}
\end{table}

\subsection{PWA - Results}

The exotic $1^{-+} f_{1}\pi^{-}$ contribution is only observed in positive reflectivity waves, indicating that the process is mediated by exchange of natural parity Reggeons, most likely $\rho(770)$ or $f_{2}(1270)$/Pomeron.  This result agrees with previous observations~\cite{Lee,ToddThesis}.  The $1^{-+}1^{+} f_{1}\pi^{-} S$ wave shows significant strength at $1.709 \ \gev$, which could be interpreted as the $\pi_{1}(1600)$, even though this mass is somewhat higher than in previous analyses~\cite{PiPiPi,EtaPrPi}.  According to predictions from the flux tube model the decay of a $J^{PC} = 1^{-+}$ hybrid into $f_{1}(1285)\pi$ should be dominant over the decay mode $\eta'(958)\pi$~\cite{HybridPhen,ClosePage}.  Comparing the present analysis of the $f_{1}(1285)\pi^{-}$ decay mode with a previous analysis of the $\eta'(958)\pi^{-}$ decay mode from the same raw data set~\cite{EtaPrPi}, we obtain the ratio
\begin{equation}\label{eq:Ratio1}
R_{1} = \frac{\text{\it BR} \ [\pi_{1}(1600) \rightarrow f_{1}(1285)\pi]}{\text{\it BR} \ [\pi_{1}(1600) \rightarrow \eta'(958)\pi]} = 3.80 \pm 0.78
\end{equation}
where the error is statistical only.  $R_{1}$ was calculated assuming that the $\pi_{1}(1600)$ reported in the $\eta'(958)\pi^{-}$ final state~\cite{EtaPrPi} is the same resonance as the low-mass state reported here in the $f_{1}\pi^{-}$ decay mode.  Since slightly different selection criteria were imposed in the two analyses, the number of $1^{-+}$ events in the present analysis were normalized by the ratio of $\eta'(958)\pi^{-}$ events in the data set from Ref.~\cite{EtaPrPi} to the observed number of $\eta'(958)\pi^{-}$ events in the present data set, prior to the mass-cut shown in Fig.~\ref{fig:ppemass}.  Table~\ref{TableComparison} lists the numbers used to determine this ratio, where the number of $1^{-+}$ events is an integration over the width of the respective Breit-Wigner distributions (see Fig.~2c in Ref.~\cite{EtaPrPi} and Fig.~\ref{fig:PWAresults}c in this work).

The observed distribution of the total $J^{PC}=1^{-+}$ strength from this analysis is also in good agreement with the results from the E818 analysis of $\pi^{-}p \rightarrow K^{+}\bar{K}^{0}\pi^{-}\pi^{-}$~\cite{Lee} and from the previous analysis of the $3\pi\eta$ final state from E852~\cite{ToddThesis}. However, the strong coupling of the low mass state to $\eta(1295)\pi^{-}$ reported in Ref.~\cite{Lee} was not observed in the two E852 analyses.  In fact the $1^{-+} \eta(1295)\pi^{-}$ waves were omitted from the fit in this work since they were very small and showed no indication of resonance behavior.  The ratio
\begin{equation}\label{eq:Ratio2}
R_{2} = \frac{\text{\it BR} \ [\pi_{1}(1600) \rightarrow f_{1}(1285)\pi]}{\text{\it BR} \ [\pi_{1}(1600) \rightarrow \eta(1295)\pi]}
\end{equation}
which is predicted to be 2.5 if the $\pi_{1}(1600)$ is a hybrid~\cite{HybridPhen}, is therefore probably quite large. Furthermore, in the theoretical estimate $R_{2}$ depends strongly on the mass of the $\pi_{1}$ and changes rapidly as the mass increases from $1.6 \ \gev$ to $1.7 \ \gev$~\cite{HybridPhen}. The low mass of this state, the value of $R_{1}$, and the uncertainty in the ratio $R_{2}$ (both experimental and theoretical) make an interpretation of the $\pi_{1}(1600)$ as a flux-tube hybrid questionable at this point. Further clarification is expected from the analysis of the $b_{1}(1235)\pi^{-}$ final state, which is currently underway~\cite{B1PiThesis}.

Because of its mass and also since it has not been seen in any of the final states $\eta\pi$, $\eta'\pi$, or $\rho\pi$, the second pole in the $1^{-+}1^{+} f_{1}\pi^{-} S$ wave at $2.001 \ \gev$ is an excellent hybrid-meson candidate. It has also been observed in the other predicted strong decay mode $b_{1}(1235)\pi^{-}$ at about the same mass~\cite{B1PiThesis}.

In the $1^{++}0^{+} f_{1}\pi^{-} P$ wave the previously observed $a_{1}(1700)$ is seen together with a higher lying state at $2.096 \ \gev$. Since the $a_{1}(1700)$ has been interpreted as the first radial excitation of the $a_{1}(1260)$~\cite{higherQuarkonia}, the $a_{1}(2096)$ emerges as a candidate for the second radial excitation or a hybrid meson with regular quantum numbers ({\em non-exotic} hybrid).  The hybrid interpretation is supported by the observation of a strong decay of the $a_{1}(2096)$ into $a_{1}(1260)\eta$ in this analysis. In order to determine the relative decay ratio between these two modes the $a_{1}(1260)\eta$ intensity was fitted to a Breit-Wigner with the mass and width fixed at the values found in the fit of the $1^{++}0^{+}f_{1}(1285)\pi$ wave (see Fig.~\ref{fig:a12eta}a).
The ratio
\begin{equation}\label{eq:Ratio3}
R_{3} = \frac{\text{\it BR} \ [a_{1}(2096) \rightarrow f_{1}(1285)\pi]}{\text{\it BR} \ [a_{1}(2096) \rightarrow a_{1}(1260)\eta]} =  3.18 \pm 0.64
\end{equation}
which is predicted to be 3~\cite{HybridPhen} if the $a_{1}(2096)$ is a hybrid, was determined assuming that $\rho\pi$ is the dominant decay of the $a_{1}(1260)$.  The error quoted for $R_{3}$ in (\ref{eq:Ratio3}) is statistical only.

Lastly, 3 poles were needed to describe the very complex $2^{-+}0^{+} f_{1}\pi^{-} D$ spectrum. The distributions show a weak signal for the $\pi_{2}(1670)$, together with two poles at $2.003 \ \gev$ and $2.46 \ \gev$.  A $\pi_{2}$ state at high mass has been observed previously~\cite{PDG,Anisovich1}, but in the present fit the state at $2.46 \ \gev$ is extremely broad and the data span only the low-mass side of the peak. A likely interpretation is that a great deal of strength in this wave may be associated with background, which can be fitted to a broad Breit-Wigner shape. The lower mass state at $2.003 \ \gev$ confirms the observation of a $2^{-+}$ isovector state at $1.9 \ \gev$, seen earlier in $a_{2}(1320)\eta$ decays~\cite{Anisovich2,Eugenio_pi2}. The $a_{2}(1320)\eta$ decay mode of the $\pi_{2}(1900)$ was also observed in this analysis and the result is shown in Fig.~\ref{fig:a12eta}b.  The $a_{2}\eta$ intensity was fitted to a Breit-Wigner shape with the mass and width fixed at the values from the $2^{-+}0^{+}f_{1}\pi$ fit.
The ratio $R_{4}$ was found to be
\begin{equation}\label{eq:Ratio4}
R_{4} = \frac{\text{\it BR} \ [\pi_{2}(1900) \rightarrow a_{2}(1320)\eta]}{\text{\it BR} \ [\pi_{2}(1900) \rightarrow f_{1}(1285)\pi]} =22.7 \pm 7.3
\end{equation}
which agrees very well with the flux-tube model prediction of $R_{4} = 23$ for the decay of a $2^{-+}$ hybrid meson at this mass~\cite{HybridPhen}.  Again, the error in (\ref{eq:Ratio4}) is statistical only.

% TABLE WITH EVENTS FROM THE COMPARISON OF F1PI AND ET'PI 
\begin{table}%[H] add [H] placement to break table across pages
\caption{\label{TableComparison}Comparison of number of $\eta'(958)\pi^{-}$ events and events in the $\pi_{1}(1600)$ peak for the analyses of $3\pi\eta$ (this work) and $\eta'(958)\pi^{-}$~\cite{EtaPrPi} from the same initial data set collected under the same trigger conditions.}
\begin{ruledtabular}
\begin{tabular}{c c c}
 &  $N(\eta'\pi^{-})$  &  $N(\pi_{1})$  \\ \hline
$3\pi\eta$ data set  &  $5 \ 885$   &  $17 \ 619 \pm 3 \ 452$  \\
$\eta'(958)\pi^{-}$ data set  &  $6 \ 040$  &  $8 \ 755 \pm 525$  \\
\end{tabular}
\end{ruledtabular}
\end{table}

% Conclusions
% ===========

\section{Conclusions}
We have performed a mass-dependent partial wave analysis of $68 \ 900$ events of the type $\pi^{-}p \rightarrow \eta\pi^{+}\pi^{-}\pi^{-} p$. Evidence is found for two resonances in the $J^{PC}m^{\epsilon}=1^{-+}1^{+}$ exotic wave in the decay mode $f_{1}(1285)\pi^{-}$, which is predicted by the flux tube model to be one of the favored decay channels for hybrid mesons with these quantum numbers.  The states are produced only in positive reflectivity, consistent with $\rho(770)$ or $f_{2}(1270)$/Pomeron exchange.  The low mass state with $M = 1.709 \ \gev$ and $\Gamma = 0.403 \ \gev$ may be identified with the previously observed $\pi_{1}(1600)$.  The second state with $M = 2.001 \ \gev$ and $\Gamma = 0.333 \ \gev$ has a mass that is in accord with predictions for a hybrid meson.  Based on its decay properties the $\pi_{1}(2000)$ must be considered a hybrid candidate at this time.  Interpretation of the $\pi_{1}(1600)$ based on existing data and model predictions is problematic.  Further clarification on the nature of both these states are expected from the results of the analysis of the $\omega\pi\pi$ final state.

In the $J^{PC}m^{\epsilon}=1^{++}0^{+}$ wave evidence is found for the $a_{1}(1700)$ in the $f_{1}(1285)\pi^{-}$ decay mode.  A second state at $2.096 \ \gev$ was observed in both the $f_{1}(1285)\pi^{-}$ and the $a_{1}^{-}(1260)\eta$ decay channels.  The mass of the higher mass state and the relative decay fraction of its two observed decay modes are in very good agreement with predictions for a non-exotic hybrid from the flux tube model.  

The $J^{PC}m^{\epsilon}=2^{-+}0^{+}$ wave shows weak production of the $\pi_{2}(1670)$ decaying into $f_{1}(1285)\pi^{-}$. Strong evidence is found for the production of the non-exotic hybrid candidate $\pi_{2}(1900)$ in the previously observed $a_{2}^{-}(1320)\eta$ channel, as well as in the $f_{1}(1285)\pi^{-}$ decay mode.  The mass and decay properties of this state are consistent with flux-tube model predictions for a non-exotic hybrid.

\begin{acknowledgments}
This research work was supported in part by the US Department of Energy, the US National Science Foundation and the Russian State Committee for Science and Technology. The authors also would like to thank the members of the Brookhaven MPS staff for their support in running the experiment.
\end{acknowledgments}

% Bibliography
\bibliography{paper}

\end{document}